# Present Velocity and Acceleration in Tide Gauge Records Characterised by a Quasi-60 years Periodic Oscillation

Thomas Watson, Bell Post Hill, VIC, Australia, E-mail: ttsw@bigpond.com

Alberto Boretti, Ballarat, VIC, Australia, E-mail: a_boretti@yahoo.com

**Abstract**

The paper describing sea level rise oscillations at Cape Hatteras, USA by Parker [1] has opened the discussion regarding if the velocity in a tide gauge record characterized by a quasi-60-year multi-decadal oscillation can be computed by linear fitting of 30 years of data in two ad-hoc selected times and if acceleration can then be inferred by comparing these two values as proposed by Sallenger [2], or if this comparison is meaningless in that the 60-year time window is the minimum amount of time needed to evaluate the velocity in a record characterized by a quasi-60-year multi-decadal oscillation and the acceleration has then to be computed as the time derivative of this velocity as suggested by Parker [1,3]. For the specific case of The Battery, NY, it is shown here that the 60-year time window is the minimum time length needed to compute a velocity, and both the 60-year windows and the all data velocities are free of any acceleration at the present time. The 30-year time window velocity of 2009 is not representative of the present sea level rise and the comparison of the 30-year time window velocity of 2009 and 1979 near a peak and a valley, respectively, of the 60-year multidecadal oscillation to claim a present acceleration has no scientific background.

**The Battery, NY tide gauge**

Even if Sallenger et al. [2] forgot the existence of a quasi-60-year periodicity in the tide gauge results along the North Atlantic coast of the United States, it is very well known that all climate parameters oscillate periodically with a quasi-60-year periodicity [4], and if sea levels do not follow temperatures only when it is of interest to the IPCC, then possibly sea levels also oscillate with this periodicity, as explicitly claimed in [3, 5]. Therefore, a proper procedure to compute the present sea level velocity and acceleration must account for the multi-decadal variability.

The least squares method is used here to calculate a straight line that best fits the data within a time window and return the window's sea level velocity (SLR). The dependent y-values are the monthly average sea levels and the independent x-values are the time in years. The calculations for $SLR_{j,k}$ is based on the formula:

$$SLR_{j,k} = \frac{\sum_{i=j}^{k}(x-x')\cdot(y-y')}{\sum_{i=j}^{k}(x-x')^2} \quad (1)$$

In this equation $x'$ and $y'$ are the sample means and $j$ and $k$ are the indices of the first and last record of the measured distribution considered for the SLR estimation. At a certain time $x_k$, $x_j$ is taken as $(x_k-30)$ to compute the $SLR_{30}$, $(x_k-60)$ when computing the $SLR_{60}$, or as $x_1$ when computing the $SLR_A$ over all years of the record. This way, from a measured distribution $x_i,y_i$ for $i=1,N$, it is possible to estimate the time histories of $SLR_{30}$, $SLR_{60}$, and $SLR_A$.

Providing that more than 60–70 years of continuously recorded data, without any quality issues, are available in a given location, the $SLR_{A,k}$ usually returns a reasonable estimation of the velocity of sea level change at the present time $x_k$ and the acceleration may then be computed as

$$SLA_k = \frac{SLR_{A,k} - SLR_{A,k-1}}{x_k - x_{k-1}} \quad (2)$$

This conventional velocity and acceleration might clearly oscillate, and their time history, rather than a single value, is of interest.

In a case with non-accelerating tide gauge records as the norm, $SLR_{1,N}$ returns the present sea level rise, and the graphs of $SLR_{j,k}$ and $SLA_k$ are only helpful to confirm the lack of any acceleration. This approach would reveal a tide gauge behaviour similar to that produced from reconstructed global mean sea level (GMSL) for 1880 to 2009, as described in Church and White [7], and would confirm the presence of an acceleration in the form of constantly rising $SLR_{j,k}$ and constantly positive $SLA_k$. However, at this stage, different mathematics would be needed to compute the present velocity and acceleration.

This simple analysis is applied here to the tide gauge data for The Battery, NY (data from [6]), one of the locations that, according to Sallenger [2], is a hot spot of positive acceleration or, according to Parker [1,3], is a location with zero acceleration. The Battery, NY (PSMSL Station ID: 12, Latitude: 40.7, Longitude: -74.013333) has a data time span from 1856 to 2011 and the data 90% complete. This record is actually composed of two records, the first 1856–1879 with one year of missed data in 1861 (23 years of data, 95% complete), and the second 1893–2012 with only a few missing months (119 years of data, ~100% complete). Considering that:

- There is some doubt regarding the data quality recorded from 1856 to 1879 (this first data oscillates around a trend line with a slope of 1.10 mm/year while the later data oscillates around a trend line of 2.80–3.10 mm/year).

- Filling the gap from 1879 to 1893 is a difficult procedure that may introduce major uncertainties.

- The available record from 1893 to 2012 satisfies all the quality and length requirements for computing the present velocity and acceleration.

We do consider here only the latter data.

Figure 1 presents the sea level velocity SLR computed with 30 and 60 years of data, and all data at any time and sea level acceleration SLA computed from all SLR data. The only conclusions that may be reasonably inferred from this picture are the following:

- The $SLR_{30}$ is not representative of the long-term velocity and largely oscillates.

- The $SLR_{60}$ and $SLR_A$ also oscillate, but the amplitude is reduced, especially in the $SLR_A$.

- The present $SLR_{30}$ is certainly larger than the $SLR_{30}$ of 30 years ago, but it is also smaller than the $SLR_{30}$ of 60 years ago.

- The present $SLR_A$ is 2.98 mm/year, while the present $SLR_{30}$ and $SLR_{60}$ are 4.48 and 3.02 mm/year, respectively.

- The present $SLR_{60}$ and $SLR_A$ have values smaller than previously measured values.

- The only reasonable conclusion that may be inferred from the present tide gauge results is that the present sea level velocity is 2.98 mm/year and there is no detectable acceleration component.

To clarify what this analysis would have produced in the case of an accelerating record, Figure 2 presents the results from a similar treatment of the reconstructed GMSL of [8]. The SLR60 is clearly always increasing and the $SLR_A$ is always positive. This is not a real measurement but a computation, not supported by any individual tide gauge of sufficient quality and length. When a tide gauge of good quality and length shows the behaviour of Figure 2, then we will possibly propose a different mathematics to better approximate the physical behaviour. For now, the basic analysis we use suffices to prove the inconvenient truth: Ocean Levels are not accelerating in Australia or around the world [9].

A comment in general should not present novel material that has not passed yet the peer review, and this comment uses as supporting material public domain information that has been previously analysed or will be analysed in other papers by the authors.

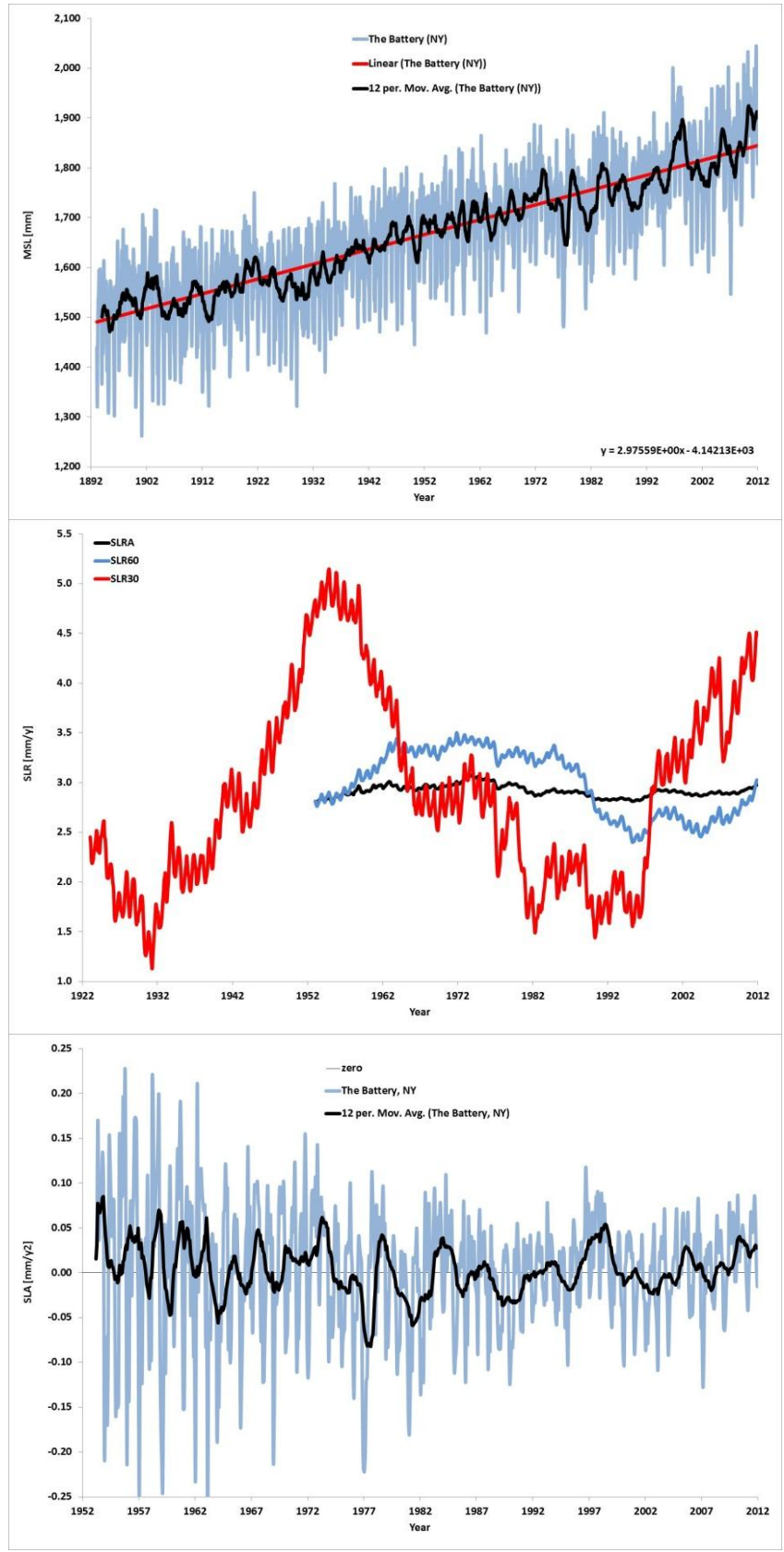

Figure 1 - Measured monthly mean sea level (MSL) data for The Battery, NY (from [6]), top. Time span of data is from 1893 to 2012. Sea level velocity (SLR) computed with 30 and 60 years of data and all data at any time, middle. Sea level acceleration (SLA) computed from all SLR data, bottom.

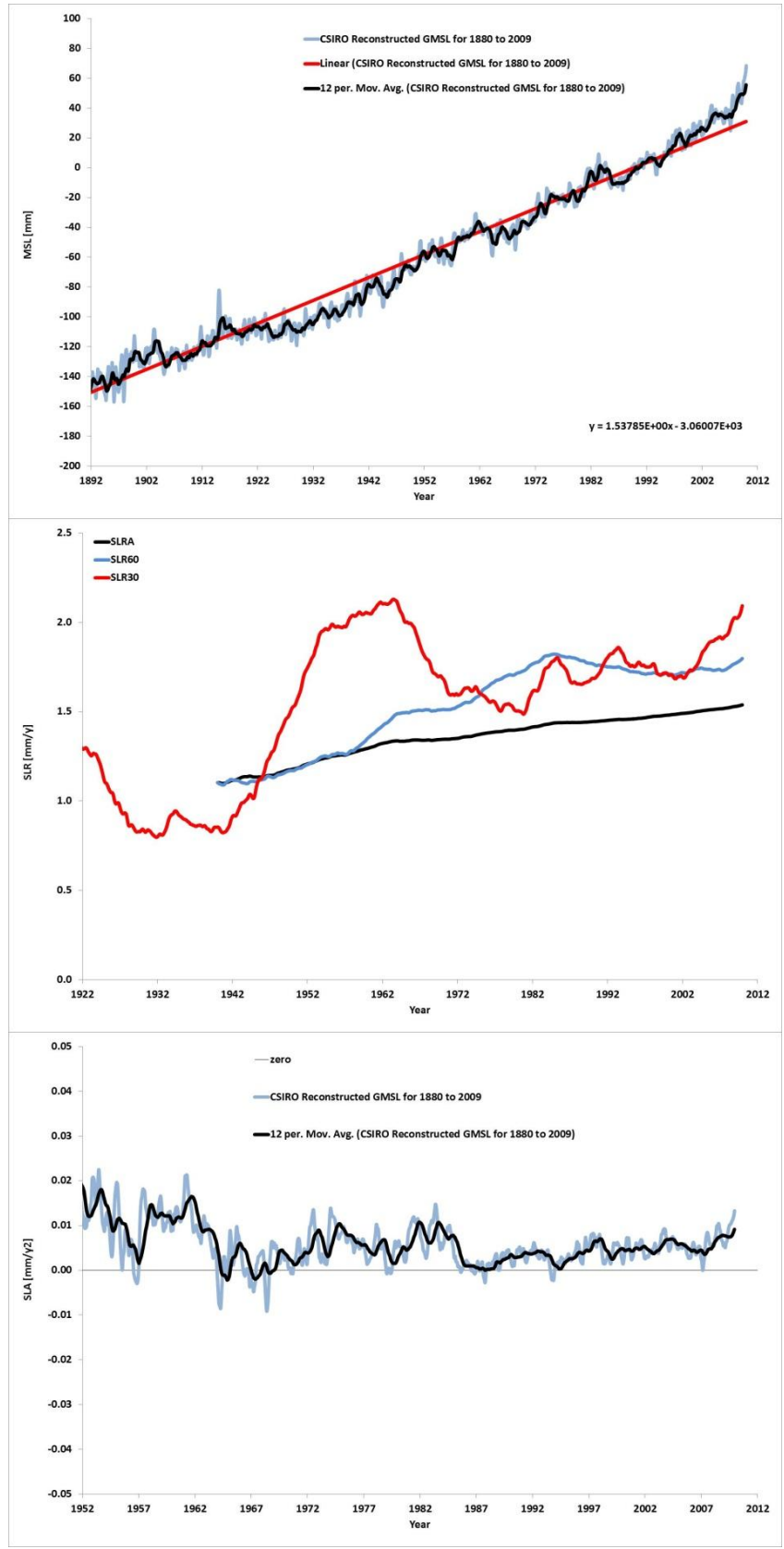

Figure 2 - Computed monthly global MSL data (from [8]), top. Time span of data is from 1880 to 2009. Sea level velocity (SLR) computed with 30 and 60 years of data and all data at any time, middle. Sea level acceleration (SLA) computed from all SLR data, bottom.